\documentstyle[12pt,moriondsissa,psfig]{article}
\def\lta{\mathrel{\mathchoice {\vcenter{\offinterlineskip\halign{\hfil
$\displaystyle##$\hfil\cr<\cr\sim\cr}}}
{\vcenter{\offinterlineskip\halign{\hfil$\textstyle##$\hfil\cr
<\cr\sim\cr}}}
{\vcenter{\offinterlineskip\halign{\hfil$\scriptstyle##$\hfil\cr
<\cr\sim\cr}}}
{\vcenter{\offinterlineskip\halign{\hfil$\scriptscriptstyle##$\hfil\cr
<\cr\sim\cr}}}}}
\def\gta{\mathrel{\mathchoice {\vcenter{\offinterlineskip\halign{\hfil
$\displaystyle##$\hfil\cr>\cr\sim\cr}}}
{\vcenter{\offinterlineskip\halign{\hfil$\textstyle##$\hfil\cr
>\cr\sim\cr}}}
{\vcenter{\offinterlineskip\halign{\hfil$\scriptstyle##$\hfil\cr
>\cr\sim\cr}}}
{\vcenter{\offinterlineskip\halign{\hfil$\scriptscriptstyle##$\hfil\cr
>\cr\sim\cr}}}}}

\newdimen\digitwidth
\setbox0=\hbox{\rm0}
\digitwidth=\wd0
\catcode`@=\active
\def@{\kern\digitwidth}

\begin{document}
{\small \sl \noindent Invited review to appear in XVIIth Moriond Astrophysics
Meeting: Extragalactic Astronomy in the INFRARED, les Arcs, France,
March 1997, ed. G.A. Mamon, T.X. Thuan \& J. Tran Thanh Van (Paris:
Fronti\`eres)} 

\vspace{-2\baselineskip}
\heading{Galaxies and Cosmology with DENIS}

\author{G.A. Mamon $^{1,2}$, M. Tricottet $^1$, W. Bonin $^{1,3}$, V. Banchet
$^{1,4}$,  } 
{$^{1}$ Institut d'Astrophysique, Paris, France.}  
{$^{2}$ DAEC, Observatoire de Paris, Meudon, France.}
{$^3$ Universit\'e Paris XI, Orsay, France.}
{$^4$ CYBEL, SA, Paris, France.}
\begin{moriondabstract}

The DENIS survey is currently imaging 21334 deg$^2$ of the mainly southern
sky in the $IJK$ and the observations are expected to go on until mid 2000.
The expectations for extragalactic and cosmological research are outlined,
including a quantitative assessment of the effects of recent star formation
on the measured fluxes of galaxies.
The galaxy extraction is much improved with the modeling of the PSF across
the $12'\times12'$ frames and the reliability of star/galaxy separation
(currently based upon a combination of classical and neural-network based
methods) is
measured from visual inspection to be $>90\%$ at $I = 16$.
The $I$ band counts follow the high bright-end
normalization and the $J$ differential 
counts follow $N(J) \simeq 11 \times {\rm dex}
[0.6 \,(J-14)]\rm\,deg^{-2}\,mag^{-1}$ and are expected to be complete,
reliable and photometrically 
accurate  ($\Delta m < 0.1$) for $J \leq 14$.

\end{moriondabstract}

\section{Cosmology and Near-Infrared selection}

Workers in the Infrared bands know that the Near-Infrared (NIR)
is 4 ($J$) to 10
($K$) times less affected by extinction from dust than is the $V$
band \cite{CCM89}.
This has two advantages: 1) A nearly-full view of the Universe, even behind
the Zone of Avoidance, and 2) A view of external galaxies unaffected by their
internal extinction by dust.

The second advantage of the Near-Infrared is its smaller bias to recent star
formation, in comparison with bluer bands and also to the mid and far IR, whose
emission arises from thermal dust created and heated by young stars.

\begin{figure}[ht]
\centerline{\psfig{file=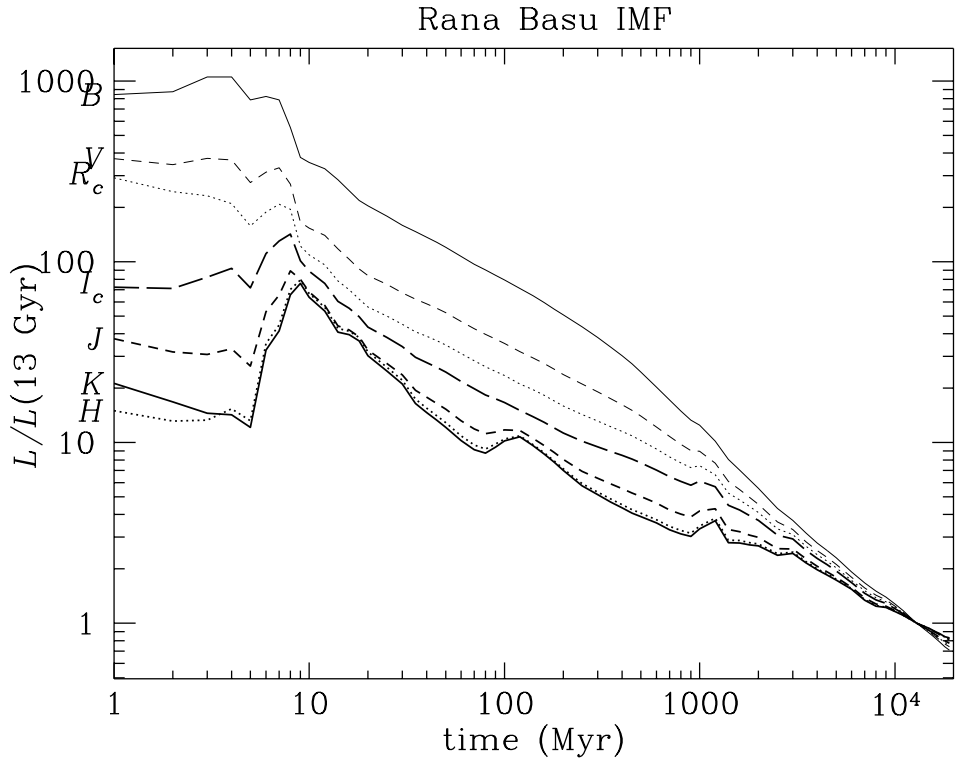,height=10cm,angle=-90}}
\caption{Time evolution of broad band luminosities (Johnson $BVJHK$ and
Cousins $RI$) of stellar burst for
stellar population with a Rana \& Basu \cite{RB92} IMF, using the 
{\sl PEGASE\/}
\cite{FR97} spectral
evolution model including nebular emission (which only affects the
luminosities within 10 Myr from the burst).}
\label{Lratiosvst}
\end{figure}

The lack of sensitivity to recent star formation can be quantified as follows.
Figure \ref{Lratiosvst} shows the evolution of broad-band luminosities,
normalized to that of a 13 Gyr old stellar population for different color
bands.
The evolution is strongest in the $B$ band (and even more so in $U$ or
far-UV, neither of which are plotted in Figure \ref{Lratiosvst}) and 
weakest in the $H$ and $K$ bands.
Nevertheless, there is a burst of emission in $H$ and $K$ roughly 10 Myr
after the starburst, when the massive stars evolve off the main-sequence to
the giant branch.
Table
\ref{SFsensitivity} below shows that the enhancement of broad band flux in the
$B$ band due to star formation occurring 1 Myr ago, relative to the $B$ flux
of an old stellar population of the same mass and IMF, is 40 times greater than
the analogous enhancement of the $K$ band flux.

\begin{table}[ht]
\begin{center}
\caption{Broad-band luminosity enhancement relative to old (13 Gyr) population
(normalized to the analogous $K$ band relative enhancement)}
\begin{tabular}{|lccccccc|}
\hline
Age & $B$ & $V$ & $R_c$ & $I_c$ & $J$ & $H$ & $K$ \\
\hline
1 Myr & 40@$\,$ & 18@$\,$ & 14@$\,$ & 3.4 & 1.8@ & 0.7@ & 1 \\
10 Myr & @5.6 & @2.4 & @1.7 & 1.4 & 1.06 & 1.05 & 1 \\
100 Myr & @7.8 & @3.5 & @2.3 & 1.6 & 1.15 & 1.03 & 1 \\
1 Gyr & @3.7 & @2.7 & @2.4 & 1.9 & 1.26 & 1.04 & 1 \\
2 Gyr & @2.1 & @1.7 & @1.5 & 1.4 & 1.11 & 1.02 & 1 \\
\hline
\end{tabular}
\end{center}
\label{SFsensitivity}
\end{table}

One interesting point is that the $H$ band is to within a few percent as
insensitive to recent star formation as the $K$ band, and
even less sensitive by up to 30\% for very recent ($< 5
\,\rm Myr$) star
formation, with the inclusion of nebular emission in the adopted 
{\sl PEGASE\/} \cite{FR97} spectral evolution model, which is stronger in
$K$ because of free-free recombination enhancing the continuum at longer
wavelengths.
Another interesting point is that the $J$ band at 1.25 $\mu$ is almost as
insensitive to recent star formation as the $K$ band, to within
typically 20\% or less, except for bursts younger than 8 Myr.
The $I_c$ band is considerably more sensitive to recent star formation than
$J$ (in Figure \ref{Lratiosvst} and Table \ref{SFsensitivity},
$I_c$ is typically
three times more distant from $K$, logarithmically, than is $J$
from $K$).

\section{The DENIS survey}

In December 1995, the DENIS (DEep Near Infrared Southern Sky
Survey \cite{E97}) began
imaging the southern sky, delimited by $-88^\circ \leq \delta \leq +2^\circ$,
in the $I_c$ ($0.8\,\mu$), $J$, and $K_s$ ($K$-short at $2.15\,\mu$)
wavebands.
Its characteristics are given in Table \ref{DENISpties} below.  The data
processing is done at the IAP (Paris) and Sterrewacht Leiden (Netherlands).
The American 2MASS project \cite{S97} will have a similar scope, with
slightly more sensitive $J$ and considerably more sensitive $K$ channels, and
an $H$ channel instead of 
our $I$ channel.
\begin{table}[ht]
\begin{center}
\caption{Characteristics of the DENIS survey (adapted from \cite{E97})}
\begin{tabular}{|ll|}
\hline
Institutions	\dotfill&	Paris-Meudon, IAP, Leiden, Grenoble,
Frascati, IAC,\\
	& Innsbr\"uck, Vienna, Lyon, Heidelberg, Budapest,\\
	& Montpellier, Besan\c{c}on, S\~ao Paolo\\
Declinations	\dotfill&	$-88^\circ \leq \delta \leq +2^\circ$ \\
Telescope	\dotfill&	ESO 1m (fully dedicated)\\
Color bands	\dotfill&	$I\, (0.8\,\mu{\rm m})$, $J\, (1.25\,\mu{\rm m})$, $K_s
\,(2.15\,\mu{\rm m})$ \\
Detectors	\dotfill&	{\small Tektronix CCD $1024^2\, (I)$,
Rockwell NICMOS-3 $256^2\, (JK)$} \\
Pixel size	\dotfill&	$1''$ ($I$), $3''$ ($JK$ dithered at $1''$)\\
Quantum efficiency	\dotfill&0.65 ($I$), 0.8 ($J$), 0.61 ($K$) \\
Exposure time	\dotfill&	$9\,{\rm s}\ (I)$, $8.8\times 1\,{\rm s}\ (JK)$ \\
Read-out noise	\dotfill&	$7\,e^-\ (I)$, $40\,e^-\ (JK)$ \\
Observing mode	\dotfill&	Stop \& stare, concurrent 3-channel with dichroics\\
Scan geometry	\dotfill&	$12'$ (RA) $\times 30^\circ$ (Dec) \\
Limiting magnitude (pt source $5\sigma$) \dotfill& 18.0 ($I$), 16.0 ($J$), 13.5
($K$) \\
Saturation magnitude \dotfill& 10.0 ($I$), 8.0 ($J$), 6.5 ($K$) \\
Survey period	\dotfill&	December 1995 -- mid 2000 \\
Data	\dotfill&	4000 GBytes (primary)	\\
Cost		\dotfill&	\$3 million	\\
\hline
\end{tabular}
\end{center}
\label{DENISpties}
\end{table}

\section{Extragalactic research and cosmology}
\label{science}

A variety of scientific questions concerning extragalactic science and
cosmology will be answered with the catalogs coming out of the DENIS and
2MASS databases:
\begin{description}
\item [Statistics of NIR properties of galaxies:]
DENIS and 2MASS will provide the first very large galaxy databases with NIR
photometry, which should 
help for distance estimates of spiral galaxies with known
velocity width (see \cite{V97}).
Surface photometry of the brighter galaxies will be correlated with color as
well as optical morphology, and with rotation velocity of known spirals
\cite{V97} and used for distance estimates for ellipticals.
\item [Cross-identification with other wavelengths:]
There will of course be plenty of cross-ident\-ification of DENIS and 2MASS
galaxies with samples at other wavelengths, such as optical galaxy samples
(see \cite{V97,SKMR97}), IRAS galaxies, quasars,
radio-galaxies, galaxies found in blind HI surveys, etc.
The NIR properties (mainly their location in color-color diagrams) of such
objects will be targeted for discovering new large samples of such objects.
There will naturally be followups at non-NIR wavelengths of DENIS and 2MASS
galaxies (see \cite{T97}).
\item [Galaxy counts:] From its $I$-band galaxy counts, DENIS should settle,
once and for all, the debate on the bright-end of the galaxy counts, where
first estimates \cite{Maddox90} saw a depletion relative to the extrapolation
of the faint-end counts, while later work \cite{BD97} disputed this.
At issue are the questions of galaxy evolution and 
whether the environment of the Local Group is
underdense on very large scales ($z \lta 0.1$). 
\item [Zone of avoidance:]
Studying galaxies behind the Galactic Plane has two main applications (see 
\cite{SKMR97}):
1) Mapping the large-scale distribution of galaxies in this still poorly
known region. Indeed, the Zone of Avoidance contains interesting structures
such as the largest 
large-scale concentration of matter in the local Universe, the Great
Attractor (at the intersection of the Supergalactic Plane and the Galactic
Plane \cite{KD95}) and within the Great Attractor, the Norma cluster, Abell
3627, richer 
and closer than the Coma cluster \cite{KKA3627}.
2) The fluxes and angular sizes of galaxies are affected by extinction from
dust in the Galactic Plane, and one can measure this extinction from galaxy
counts
\cite{BH82}, colors \cite{MBTK97}, and color-color diagrams \cite{SKMR97}.
\item [Structures of galaxies:]
Only a few catalogs of clusters \cite{LNCG92,DEMS94,EM95} and compact groups
\cite{PIM94} are based
upon 
automatically 
selected galaxy samples, which happen to be 
optical and photographic (hence subject to
photometric non-linearities).
Because star formation is probably enhanced by galaxy interactions, one
expects that the statistical properties of pairs, groups and clusters of
galaxies built from NIR selected galaxy catalogs will be different from those
built from optical catalogs.
DENIS and 2MASS will thus have the double advantage of using an NIR
galaxy sampled based upon linear (non-photographic) photometry.
The applications of such NIR-based samples of structures of galaxies are
numerous \cite{M94} and include understanding the dynamics of these
structures, 
their bias to projection effects, their constraints on $\Omega_0$
and the primordial density fluctuation spectrum, their use as distance
indicators, and the environmental
influences on galaxies.
\item [The large-scale structure of the Universe:]
The NIR selection and the linear photometry will also benefit the measurement
of statistics (two-point and higher-order angular correlation functions,
counts in cells, 
topological genus, etc.) of the large-scale distribution of galaxies in the
Universe.
For example, the (3D) primordial density fluctuation spectrum of galaxy
clustering can be obtained from the
two-point angular correlation function \cite{BE93} or from the 2D power
spectrum \cite{BE94}.
Moreover, by the end of DENIS and 2MASS, large-scale cosmological simulations
with gas dynamics incorporated (thanks to which galaxies are properly
identified) will provide adequate 
galaxy statistics in projection that will be compared
with those obtained from the surveys, iterating over the cosmological input
parameters of the simulations.
\end{description}

DENIS has already made contributions in stellar astronomy, with a new short
scale length (2.3 kpc) and cutoff (15 kpc) of the Galactic disk \cite{R96}
and the discovery of three
candidate brown dwarfs \cite{DF97}, one of which ($\leq 0.065\,M_\odot$ in
Corvus) has been confirmed
spectroscopically \cite{TDF97}.

\section{Galaxy extraction}

The advent of first DENIS results in stellar astronomy highlights the
difficulty of extracting galaxies from the DENIS images.
Indeed, at the star/galaxy separation limits of the $I$ band (which is the
most sensitive, see \S\ \ref{limits}), there are 3 to
5 times more stars than galaxies at high galactic latitudes \cite{LP96}, 
and up to $10^4$ times
more in the Galactic Plane (with considerably less favorable ratios at
brighter magnitudes, since galaxy counts rise faster than star counts).
Whereas much of the science outlined in section \S\ \ref{science} requires
galaxy samples that are complete and reliable (with few 
stars or optical defects misidentified as galaxies), such complete and
reliable galaxy extraction is hindered by the current weaknesses  in the
DENIS camera:
\begin{itemize}
\itemsep -3pt
\item Low sensitivity in $K$ caused by the thermal emission of the instrument
\item Undersampled PSF in $J$ and $K$ (despite the dithering mentioned in
Table \ref{DENISpties})
\item Large PSF in $I$ (typically $2''$), caused in part by defocusing, coma
and astigmatism of the camera
\item Important PSF variations across the frames (10\% to 50\%) with the same
causes plus optical misalignment
\item Elongated PSFs near the frame edges
\item Artifacts caused by readout electronics
\end{itemize}
In particular, the star/galaxy separation in the $I$ band depends
crucially on the accurate modeling of the PSF across the DENIS frames.
An r.m.s. error of 5\% on the FWHM of the PSF leads to a loss of 0.5
magnitude in the completeness limit of galaxy extraction at given reliability
of star/galaxy separation \cite{B98}.

\subsection{Pipeline}

We have worked on the images that were bias subtracted, flat-fielded,
background subtracted, and bad pixel flagged with the software written by
J. Borsenberger \cite{B97}, who has also performed absolute astrometry by
tying to a copy at the CDS in Strasbourg 
of the USNO-A1.0 \cite{M97} astrometric 
catalog obtained from PMM
scans.
One of us (G.A.M.) has implemented a preliminary galaxy
pipeline with the following 
steps:
\begin{itemize}
\itemsep -3pt
\item Cosmic ray and bad pixel filtering
\item Reading photometric zero-points and airmasses from relevant files
\item Galaxy extraction using {\sl SExtractor} \cite{BA96}, version
1.2b6a (including a neural-network based star/galaxy separation \cite{B95},
whose input parameters are 8 isophotal areas, the maximum intensity and as a
control parameter, the
FWHM of the PSF),
with detection and Kron photometry \cite{K80} 
parameters optimized from simulated images
\cite{B98}
\item Modeling of the variations of the object FWHMs across the frame 
(with a program written by M.T.)  
\item  Star/galaxy separation using the updated PSF at the object position
obtained from the modeling across the frame, using E. Bertin's {\sl
SExPlay2\/}
\end{itemize}

\subsection{Photometric accuracy}

\begin{figure}[ht]
\centerline{\psfig{file=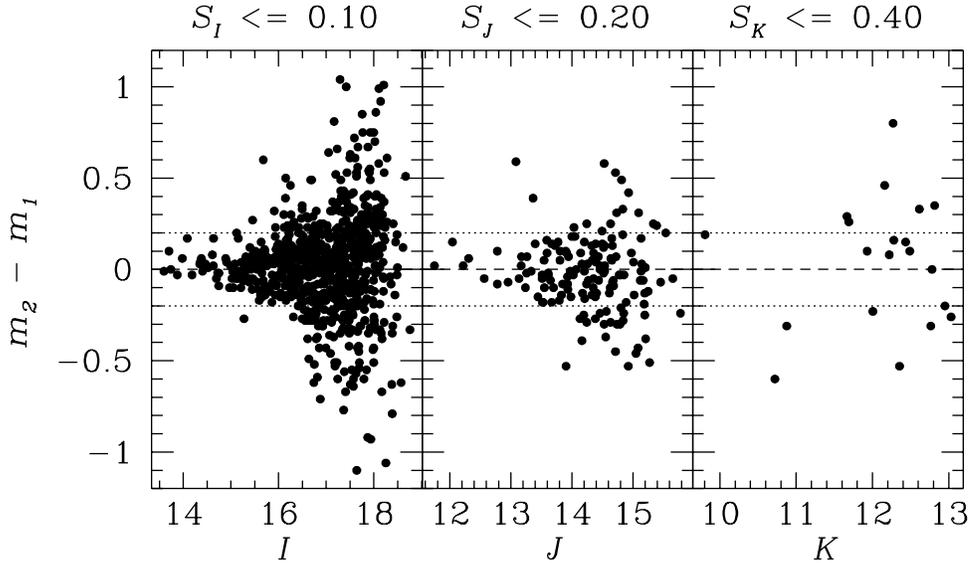,width=0.8\textwidth,angle=-90}}
\caption{Photometric difference on overlap (concordant astrometry $<2''$)
candidate galaxies on  $50\,\rm deg^2$,
discounting flagged (blends, edge) objects and objects with centers within 20
pixels from edge of frames.
The limiting neural network \cite{BA96} stellarity index (0 for extended
objects, 1 for 
stars) for galaxies is given at the top (PSF modeling was used, see \S\
\ref{sep}).
}
\label{overlaps}
\end{figure}

Photometry is an important aspect in understanding the diagnostics of galaxy
extraction and of most algorithms of classical (not based on neural networks)
star/galaxy separation. 
We report here on the first measures of photometric accuracy of DENIS
candidate galaxies, using objects
lying in the $2'$ overlaps of adjacent images.
Figure \ref{overlaps} shows the differences in photometry on overlap objects
that are likely to be galaxies.
The photometric accuracy on a given individual measurement is
$\langle m_1-m_2 \rangle_{\rm rms} / 2^{1/2}$.
We thus infer from Figure \ref{overlaps} that $\sigma_I = 0.08$ at $I = 17$,
$\sigma_J = 0.08$ at $J = 14.5$, and $\sigma_K = 0.42$ (0.31) at $K = 12$
(12.5).
These do not include systematic photometric errors caused by uncertain
zero-points, extrapolations to total magnitude, etc.
The small number of overlaps in $K$ means that the photometric accuracy in
$K$ is uncertain.

\subsection{Star galaxy separation}
\label{sep}

The optical deficiencies of the DENIS camera cause variations of the
PSF across the DENIS frames, which results in an excess of objects classified
as extended (galaxies) in regions where the PSF is larger, if no modeling of
the PSF is performed 
before star/galaxy separation \cite{MBTK97}.
One of us (M.T.) has
modeled the PSF FWHM as a two-dimensional second order polynomial
across the frame using unflagged objects with $I \leq 16.2$ and iteratively
rejecting $2.5\,\sigma$ outliers (see also \cite{B97}).

Figure \ref{svsm} shows how the neural network stellarity \cite{BA96}
seems to discriminate better
between stars and galaxies once PSF modeling is incorporated, providing a
gain of roughly one magnitude in the reliability limit (the DENIS strip 5544
used here spans galactic latitudes
from $15^\circ$ to $34^\circ$ towards the Galactic Center).

\begin{figure}[htp]
\vspace{7.5cm}
\caption{Neural network \cite{BA96} stellarity parameter (1 = stellar, 0 =
extended) 
versus magnitude for 180 frames of fairly low latitude strip 5544,
without ({\it left\/}) and with 
({\it right\/}) PSF modeling.} 
\label{svsm}
\end{figure}
\begin{figure}[htp]
\vspace{12cm}
\caption{Positions of candidate galaxies ($I \leq 16.0$ and limiting neural
network stellarity 
index for galaxies $S_I \leq 0.1$, only blending flags
permitted) on DENIS frames, stacked over DENIS
strips
5544 ({\it top\/}) and 5570 ({\it bottom\/}, $33^\circ \leq b \leq
61^\circ$), each containing 180 images. 
The {\it left\/} and {\it right\/} plots correspond to neural network
stellarities defined 
without and with PSF modeling, respectively.
Circle diameters vary as the square root of the fluxes.}
\label{xyplots}
\end{figure}
%

%

Figure \ref{xyplots} shows the positions of candidate galaxies in the $I$
band
before and after modeling the PSF variations.
Vignetting is apparent in both strips and corrected by the PSF modeling,
perhaps too much so.
Also, the upper left quadrant appears to produce an excess of galaxy
candidates before PSF modeling, which may also be present after PSF
modeling.
However,
the positions of the extended objects appear to 
lack uniform coverage of the frame
in the PSF modeled star/galaxy separation.
Hence, the PSF modeling is still imperfect and can probably be improved.
We are currently analyzing new ways of modeling the PSF for adequate
uniformity of the extracted galaxies.

One of us (W.B.) has obtained estimates of the reliability of star galaxy
separation through systematic visual inspection of 56 objects in one $6\,\rm
deg^2$ strip (5544), with $I \leq
16.0$ and who figure in the galaxy branch of the area-magnitude diagram shown
in Figure \ref{avsm}.
Borderline cases were checked visually by a second person (G.A.M.).
91\% of objects passing both classical area vs. magnitude and neural network
star/galaxy separation were confirmed as 
galaxies, while 7\% were double stars (half of
these were in an unusually dense region) and 2\% (1 object) was an optical
default (the spike of a mildly saturated star).

\begin{figure}[ht]
\vspace{10.5cm}
\caption{Area versus $I$ magnitude for DENIS strip 5544 (objects at least 20
pixels from frame edges).
{\it Circles\/} are candidate galaxies with stellarity $S_I \leq 0.10$, the
{\sl line\/} is 
a fiducial classical star/galaxy separation, and the {\it plus signs\/} above
the 
line are classified as stars by the neural network ($S_I > 0.10$).
}
\label{avsm}
\end{figure}
Similarly, we visually checked the 31 objects classified as stars by the neural
network but situated above the classical dividing line in the area-magnitude
diagram, and {\it all\/} were confirmed stars though 2 (6\%) seemed to have
faint galaxies superposed.
For example, the isophotal area of the
bright object at $I = 12$ classified as a star was seriously overestimated
because it lied near a saturated star, but this did not fool the neural
network. 

\section{Preliminary results on 50 deg$^{\bf 2}$}

There are two reasons to rely on
$I$ band star/galaxy separation when building lists of galaxies in $J$ or
$K$: 1) $I$ is  considerably more sensitive than $J$ and $K$ (except in
highly extinguished regions, $A_V \gta 6$ \cite{SKMR97}), and 2) $I$ has
better angular resolution.
Since typical $I-J$ colors are almost always bluer than 1.6 (see below), $I$
band star/galaxy 
separation will be roughly 90\% reliable (see \S\ \ref{sep} above) down to $J =
16 - 1.6 = 14.4$ and for $K \lta 13.2$ (for similar reasons), 
well beyond the completeness limits for galaxy detection
(see Table \ref{tb:limits} below).

\begin{figure}[ht]
\centerline{\psfig{file=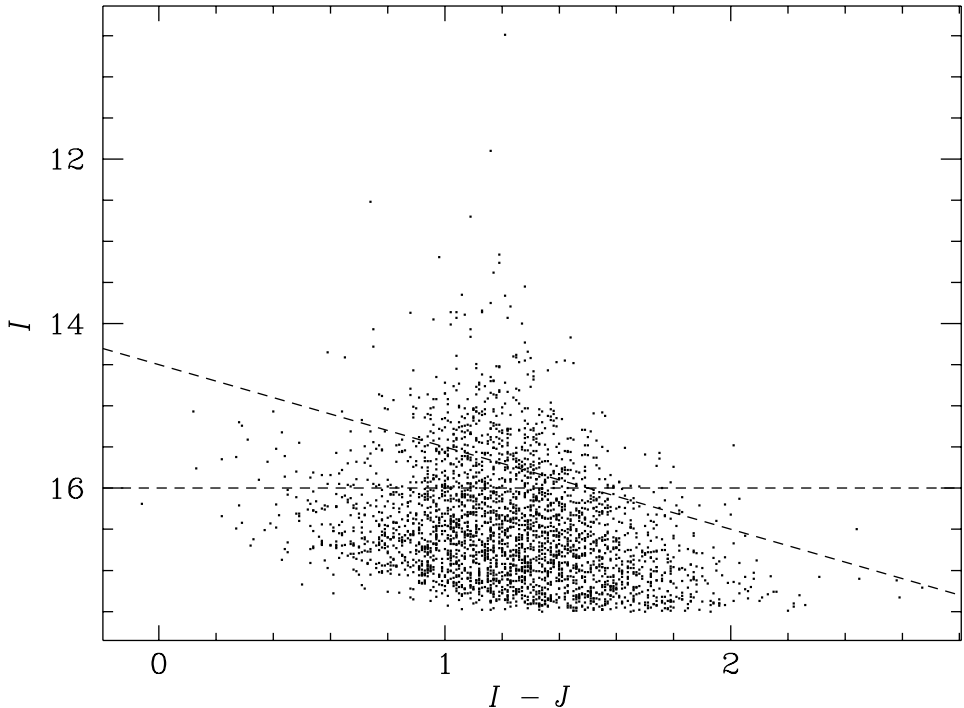,angle=-90,width=0.8\textwidth}}
\caption{Color magnitude diagram for galaxies from 8 high
latitude DENIS strips ($S_I \leq 0.1$ and isophotoal area in $I$ above
critical line of Figure \ref{avsm}, with no additional
star/galaxy separation in $J$; objects closer than 50 pixels to the frame
edges are excluded). The {\it horizontal line\/} represents $I = 16$ and
the {\it tilted line\/} represents $J = 14.5$.
}
\label{colormag}
\end{figure}

The results below are based upon $I$ band star/galaxy separation using
stellarity ($S_I \leq 0.10$) and area vs. magnitude (dividing line in Figure
\ref{avsm}).
Figure \ref{colormag} presents the $I$,$I-J$ color-magnitude diagram for 8
high galactic latitude DENIS strips.
The increased scatter at faint $I$ magnitude is caused by photometric errors.
The very blue and very red objects are probably
misclassified stars. 
In any event, only the points above the two lines in Figure \ref{colormag}
have reliable star/galaxy separation.
The galaxy locus seems to occur in the region $0.6 \leq I-J \leq 1.6$.
At $J = 14.0 \ (14.4)$, 96\% (90\%) of the galaxies have $I
\leq 16.0$, so that 
this $J$ list is over 87\% (82\%) reliable.

Note that the magnitudes in these plots are based upon Kron \cite {K80}
photometry, which 
attempts to approximate the total magnitude of galaxies.
Since, $I$ is more sensitive than $J$, the Kron aperture radius is larger in
$I$. We thus expect fixed aperture colors to appear slightly redder.

\begin{figure}[htp]
\centerline{\psfig{file=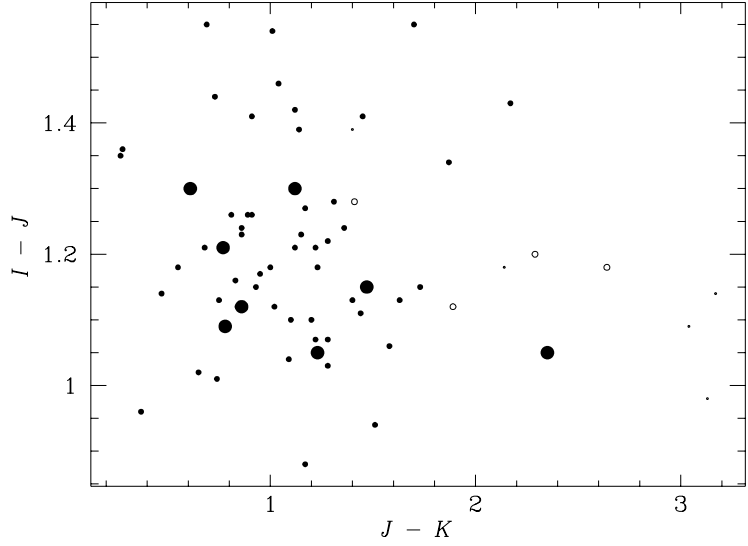,angle=-90,width=0.61\textwidth}}
\caption{Color-color diagram for galaxies (using the same star/galaxy
separation as in Figure \ref{colormag} and no additional star/galaxy
separation in $K$) in 8 high
latitude DENIS strips.
The {\it large circles\/} correspond to $I \leq 16.0$ and 
$K \leq 11.5$, the {\it small circles\/} to $I \leq 16.0$ and $K > 11.5$,
the {\it dots\/} to $I > 16$. 
The {\it solid\/} and {\it open circles\/} correspond to
$J \leq 14.4$ and $J > 14.4$, respectively.}
\label{colcol}
\end{figure}

In figure \ref{colcol}, we show the $IJK$ color-color diagram for our 8 high
latitude strips.
\begin{figure}[htp]
\centerline{\psfig{file=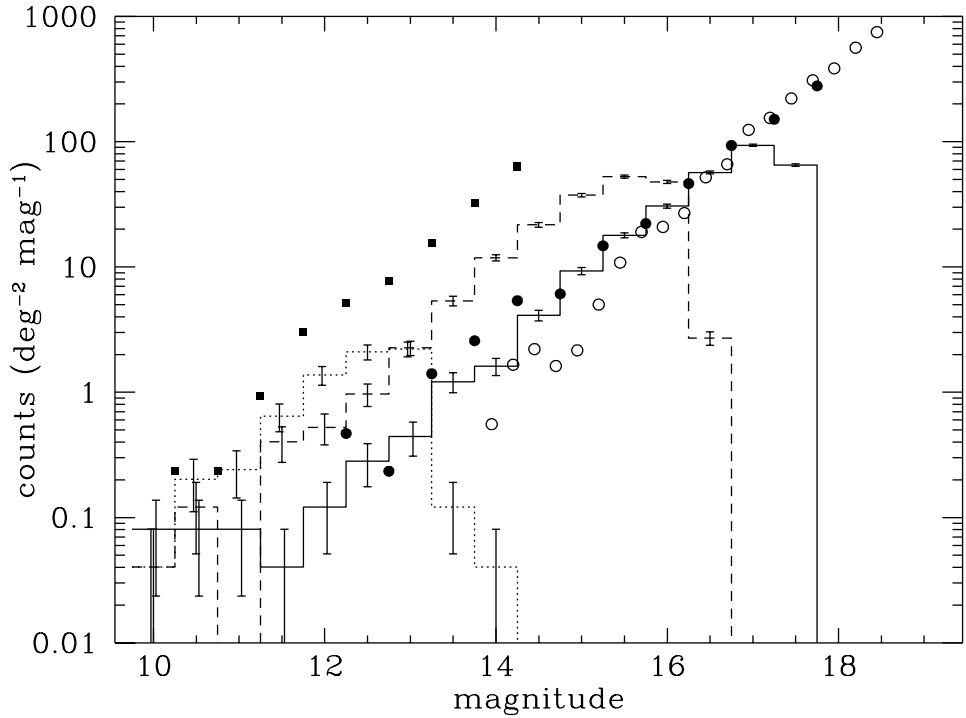,angle=-90,width=0.79\textwidth}}
\caption{DENIS differential galaxy counts (using the same star/galaxy
separation as in Figure \ref{colormag}, but with an additional
color criterion deduced from
Figure \ref{colormag}: $0.6 \leq I-J \leq 1.6$). Objects closer than 50 pixels
to the 
frame edges are excluded, for a total equivalent area of $50\,\rm deg^2$.
The {\it dotted\/} ({\it left\/}), {\it dashed\/} ({\it middle\/}), and {\it
solid\/} ({\it right\/}) {\it histograms\/} correspond to
the DENIS counts in $K$, $J$, and $I_c$, respectively.
Data points are counts from
Gardner {\it et al.\/} \cite{GSCF96} in $K$ ({\it squares\/}) and $I_c$ ({\it
filled circles\/}) and from Lidman \& Peterson \cite{LP96} in $I_c$ ({\it open
circles\/}).
}
\label{counts}
\end{figure}
%
%
The galaxies with reliable photometry ({\it large filled circles\/} in Figure
\ref{colcol}) cluster around $I-J \simeq 1.2$ and $J-K \simeq 1.1$, and an
important part of the scatter is caused by photometric errors.
Again, the colors within fixed apertures will be redder (see above).
We are in the process of visualizing the galaxies that appear unusually
red in $J-K$, but we note that
most of these are too faint for very reliable star/galaxy separation in $I$
and/or accurate photometry in $J$
or $K$.

Figure \ref{counts} shows the galaxy 
counts in $IJK$ bands for the 8 high galactic
latitude DENIS strips.
The DENIS $I$ counts follow Gardner {\it et al.\/}'s \cite{GSCF96} out to a
75\% completeness limit of $I \simeq 17.25$.
The disagreement at the bright end with Lidman \& Peterson \cite{LP96} argues
for a {\it high bright-end normalization of galaxy counts\/} (see
\cite{BD97}), although not as high as Gardner {\it et al.\/}'s.

Note that DENIS, Gardner {\it et al.\/} and Lidman \& Peterson all work with
the 
Cousins $I$ band, so no conversion was made from another $I$ filter.
Also, our survey has smaller error bars at the bright end as it covers 
4 to 5 times the solid angle of the two cited
surveys.
In comparison with Gardner {\it et al.\/}'s \cite{GSCF96} $K$ counts, the
DENIS $K$ counts become incomplete at $K \simeq 11.25$.
Figure \ref{counts} shows what we believe are the first reliable bright
galaxy counts 
in $J$, and are roughly 80\% (50\%) complete out to $J \simeq 15.25 \ (15.75)$.
The $J$ counts can be parameterized by $N(J) \simeq 11 \times {\rm dex}
[0.6 \,(J-14)]\rm\,deg^{-2}\,mag^{-1}$. 

\section{Discussion}
\label{limits}

With the plots presented above, we are now in a position to provide our
expected limits for the DENIS survey.
{}From Table \ref{tb:limits} below, we infer that star/galaxy separation is the
limiting factor in $I$ and $J$, while completeness and 
accurate (0.20 magnitude) photometry
are the limiting factors in $K$.
\begin{table}[ht]
\caption{Estimated DENIS limits from $50\,\rm deg^2$ of reduced data}
\begin{center}
\begin{tabular}{|lccc|}
\hline
	& $I_c$	& $J$ & $K$ \\
\hline
completeness ($\simeq 80\%$) & 17.25 & 15.25 & 11.25 \\
star/galaxy separation (90\% reliability, from $I$) & 16.0@\ & 14.0@ & 12.5@ \\
photometry (0.10 magnitude accuracy) & 17.4@ & 15.0@ & $<$11?@@@ \\
\hline
\end{tabular}
\end{center}
\label{tb:limits}
\end{table}
And from the counts of Figure \ref{counts}, we estimate that complete,
reliable, and photometrically accurate
DENIS catalogs will include $600\,000$, $200\,000$, and $< 8000$ galaxies at
$I < 16.0$, $J < 14.0$, and $K < 11.2$, respectively.
A cooling system has been very recently installed on the DENIS camera,
and we hope that this will improve the $K$-band sensitivity.
We also expect that better star/galaxy separation, in particular through better
modeling of the PSF variations across the frames, will increase the
reliability limit up to $I \simeq 16.5$, which will roughly double the number
of galaxies in the complete and reliable $I$ and $J$ band catalogs.

\acknowledgements{We are grateful to 
P. Fouqu\'e and the DENIS operations team for running the DENIS telescope,
J. Borsenberger for developing the
pre-extraction and astrometric calibration software necessary for the results
presented here,
E. Bertin for providing numerous updates to his {\sl SExtractor\/}
and {\sl SExPlay2\/} software used for galaxy extraction and 
star galaxy separation, and for comments on the manuscript,
and M. Fioc for providing intermediate results from his {\sl PEGASE\/} 
spectral evolution software.}


%
\end{document}